\documentclass[aps,amsmath]{revtex4}
\usepackage{graphicx}
\usepackage{pstricks}

\def\al{\alpha}
\def\be{\beta}
\def\ga{\gamma}
\def\de{\delta}
\def\ep{\epsilon}

\def\et{\eta}
\def\th{\theta}

\def\io{\iota}
\def\ka{\kappa}
\def\la{\lambda}

\def\rh{\rho}
\def\vr{\varrho}
\def\si{\sigma}

\def\ph{\phi}
\def\vp{\varphi}
\def\ch{\chi}

\def\mn{{\mu\nu}}
\def\kl{{\ka\la}}
\def\rs{{\rh\si}}

\def\prt{\partial}
\def\cl{{\cal L}}
\def\half{\tfrac12}

\newcommand{\beq}{\begin{equation}}
\newcommand{\eeq}{\end{equation}}
\newcommand{\bea}{\begin{eqnarray}}
\newcommand{\eea}{\end{eqnarray}}
\newcommand{\rf}[1]{(\ref{#1})}
\def\etal{{\it et al.}}
\def\mbf#1{\mbox{\boldmath$#1$}}

\DeclareMathOperator{\sgn}{sgn}
\def\cg#1#2{\langle #1 | #2 \rangle}
\def\bc#1#2{\left(\begin{smallmatrix} #1 \\ #2 \end{smallmatrix}\right)}

\def\yjm#1{Y_{#1}}
\def\syjm#1#2{{}_{#1}Y_{#2}}
\def\Yjm#1{{\mathcal Y}_{#1}}
\def\Yrjm#1#2{{\mathcal Y}^{#1}_{#2}}
\def\Yrjmconj#1#2{{\mathcal Y}^{#1*}_{#2}}
\def\sNrqj#1#2#3#4{{}_{#1} N^{#2#3}_{#4}}

\def\Aqjm#1#2{{A^{#1}_{#2}}}
\def\Bqjm#1#2{{B^{#1}_{#2}}}
\def\Cm#1{C_{#1}}
\def\Dr#1{D_{#1}}
\def\Fr#1{E_{#1}}
\def\Tr#1{T_{#1}}

\def\u{\text{\scalebox{0.8}{$\uparrow$}}}
\def\d{\text{\scalebox{0.8}{$\downarrow$}}}

\def\qud{{\widehat q_{\u\d}}}
\def\qpm{{\widehat q_\pm}}

\def\r{\vr}

\def\n{n}
\def\N{\mbf n}
\def\Nx{n_x}
\def\Ny{n_y}
\def\Nz{n_z}
\def\Nu{n_\u}
\def\Nd{n_\d}

\def\E{\mbf e}
\def\Ex{\E_x}
\def\Ey{\E_y}
\def\Ez{\E_z}
\def\EX{\E^x}
\def\EY{\E^y}
\def\EZ{\E^z}
\def\Eu{\E_\u}
\def\Ed{\E_\d}
\def\EU{\E^\u}
\def\ED{\E^\d}
\def\Eth{\E_\th}
\def\Eph{\E_\ph}
\def\ETH{\E^\th}
\def\EPH{\E^\ph}
\def\Er{\E_r}
\def\Ep{\E_+}
\def\Em{\E_-}
\def\Epm{\E_\pm}

\def\ER{\E^r}
\def\EP{\E^+}
\def\EM{\E^-}

\def\EMP{\E^\mp}

\def\rhat{{\hat r}}

\def\K{\mathcal K}

\def\q{q}

\def\s{s}

\def\k{k}

\def\kNdjm#1#2{k^{{\rm N}(#1)}_{#2}}

\newcounter{tc1}\newcounter{tc2}
\newcounter{tr1}\newcounter{tr2}
\newlength{\h}
\def\BT#1#2{\psset{unit=12pt,linewidth=0.5pt}%
  \setlength{\h}{#2\psunit}\setlength{\h}{0.5\h}\addtolength{\h}{-0.3\psunit}
  \begin{pspicture}[shift=-\h](#1,#2)\small%
    \setcounter{tc1}{0}\setcounter{tc2}{1}%
    \setcounter{tr1}{#2}\setcounter{tr2}{#2}\addtocounter{tr1}{-1}%
    \psline(0,0)(0,#2)(#1,#2)}
\def\ET{\end{pspicture}}
\def\BX#1{%
  \psline(\value{tc1},\value{tr1})(\value{tc2},\value{tr1})(\value{tc2},\value{tr2})%
  \rput(\value{tc1},\value{tr1}){\rput(0.5,0.5){\ensuremath{#1}}}
  \addtocounter{tc1}{1}\addtocounter{tc2}{1}}

\def\ROW{%
  \addtocounter{tr1}{-1}\addtocounter{tr2}{-1}%
  \setcounter{tc1}{0}\setcounter{tc2}{1}}

\def\S{{\mathcal S}}
\def\A{{\mathcal A}}
\def\P{{\mathcal P}}
\def\C{{\mathcal C}}
\def\Num{{\mathcal N}}

\begin{document} 

\title{Spherical-harmonic tensors}

\author{Francisco Gonzalez Ledesma$^1$ and Matthew Mewes$^2$}
\affiliation{
  $^1$Department of Physics,
  Florida State University,
  Tallahassee, Florida 32306, USA
  \\
  $^2$Physics Department,
  California Polytechnic State University,
  San Luis Obispo, California 93407, USA}

\begin{abstract}
  The connection between spherical harmonics
  and symmetric tensors is explored.
  For each spherical harmonic,
  a corresponding traceless symmetric tensor is constructed.
  These tensors are then extended to include nonzero traces,
  providing an orthonormal angular-momentum eigenbasis
  for symmetric tensors of any rank.
  The relationship between the spherical-harmonic tensors
  and spin-weighted spherical harmonics is derived.
  The results facilitate the spherical-harmonic
  expansion of a large class of tensor-valued functions.
  Several simple illustrative examples are discussed,
  and the formalism is used to derive the leading-order
  effects of violations of Lorentz invariance
  in Newtonian gravity.
\end{abstract}

\maketitle

\section{Introduction}

Spherical harmonics $\yjm{jm}$
provide an orthonormal basis
for scalar functions on the 2-sphere
and have numerous applications
in physics and related fields.
While they are commonly written in terms
of the spherical-coordinate polar angle $\th$
and azimuthal angle $\ph$,
spherical harmonics can be expressed
in terms of cartesian coordinates,
which is convenient in certain applications.
The cartesian versions involve
rank-$j$ symmetric trace-free tensors $\Yjm{jm}$
\cite{applequist,herrmann,thorne,poisson,stone}.
These form a basis for traceless tensors
and provide a link between functions on the sphere
and symmetric traceless tensors in three dimensions.

This work builds on the above understanding in several ways.
We first develop a new method for calculating
the scalar spherical harmonics $\yjm{jm}$
in terms of components of the direction unit vector
\beq
\N = \sin\th\cos\ph\,\Ex + \sin\th\sin\ph\,\Ey  + \cos\th\,\Ez \ .
\label{nhat}
\eeq
The result can be used to write the spherical harmonics
in terms of cartesian coordinates,
spherical-coordinate angles,
or any other coordinates.
We then extract
the traceless $\Yjm{jm}$ tensors
and study their properties.
These are extended to rank-$\r$ tensors $\Yrjm{\r}{jm}$
with nonzero trace,
which can be used to perform both a trace
and angular-momentum decomposition of an arbitrary tensor.
The formalism is then generalized
to spin-weighted spherical harmonics $\syjm{s}{jm}$
\cite{sYjm1,sYjm2,sYjm3}
and tensor-valued function spaces.

Spherical harmonics are eigenfunctions
of angular momentum $\mbf J = \mbf S + \mbf L$,
with eigenvalues $J^2 = j(j+1)$
and $J_z = m$,
where $m$ is limited by $|m|\leq j$.
Angular momentum is the generator for rotations,
so spherical harmonics provide a natural characterization
of the rotational properties
and direction dependence of a system.
For a scalar function $f(\N)$,
the spin $\mbf S$ is zero,
and $\mbf J$ is purely orbital angular momentum $\mbf L$,
which accounts for the functional dependence on $\N$.
The spherical decomposition 
$f(\N) = \sum_{jm} f_{jm} \yjm{jm}(\N)$
involves quantum numbers $\{j,m\}$
associated with the compatible operators $\{J^2,J_z\}=\{L^2,L_z\}$.
Each term in the expansion represents just one example
of a structure with definite $J^2$ and $J_z$.

In contrast to scalar functions,
constant tensors are pure-spin objects
with zero orbital angular momentum.
Consider, for example,
a constant traceless symmetric tensor $T$ of rank $\r$.
In this case, the total spin and total angular momentum
are both  $j=\r$.
The symmetry of $T$ implies a total of $2j+1$ independent components,
matching the number of $m$ values for fixed $j$.
It can be expanded in spin-eigenbasis tensors $\Yjm{jm}$,
$T = \sum_{m} T_{m} \Yjm{jm}$.
Each term in this expansion has the same total angular momentum
as the corresponding $\yjm{jm}$ term in the expansion of
the scalar $f(\N)$,
but in the form of spin rather than orbital angular momentum..

It is not surprising that a connection exists
between spherical harmonics $\yjm{jm}(\N)$ and
the $\Yjm{jm}$ basis tensors.
In fact,
the contraction of the $\Yjm{jm}$ tensor with the $\N$ vector $j$ times
yields a scalar function proportional to $\yjm{jm}(\N)$.
This provides a link between scalar functions
and constant tensors,
a relation that can be generalized to tensor-valued functions.
Contracting $\Yjm{jm}$ with a single $\N$ vector
gives a traceless symmetric rank-$(j-1)$ tensor function of $\N$.
This decreases the spin by one and
increases the orbital angular momentum by one,
while leaving the total angular momentum unchanged.
Subsequent contractions with $\N$
continue to convert spin angular momentum
to orbital angular momentum until we arrive
at the scalar spherical harmonics $\yjm{jm}(\N)$.
Consequently,
each $\Yjm{jm}$ generates a set of $j+1$ tensor-valued
eigenfunctions of $J^2$ and $J_z$
with different ranks.
For example,
the tensor $\Yjm{30}$ generates four different
angular-momentum eigenfunctions:
the spin-3 constant $(\Yjm{30})^{abc}$,
the spin-2 $(\Yjm{30})^{abc}\n_c$,
the spin-1 $(\Yjm{30})^{abc}\n_b\n_c$,
and the scalar $(\Yjm{30})^{abc}\n_a\n_b\n_c$.
This procedure yields a natural set of
tensor spherical harmonics of different ranks and spins.
The components of these tensors in a special helicity basis \cite{km09}
are the spin-weighted spherical harmonics
up to a normalization factor.

This paper is organized as follows.
The basic theory is given in Sec.\ \ref{theory}.
Section \ref{notation} establishes some notation and conventions.
A new expression for scalar spherical harmonics $\yjm{jm}$
in terms of the components of $\N$ is derived in
Sec.\ \ref{scalarY}.
This expression is used in Sec.\ \ref{Ytensors} to construct
the traceless rank-$j$ spherical-harmonic tensors $\Yjm{jm}$.
Section $\ref{Yrtensors}$ extends
the $\Yjm{jm}$ to rank-$\r$ tensors $\Yrjm{\r}{jm}$ with nonzero trace.
The connection between the $\Yrjm{\r}{jm}$
and spin-weighted spherical harmonics $\syjm{s}{jm}$
is derived in Sec.\ \ref{tensorY}.
Some simple illustrative examples are given in Sec.\ \ref{examples}.
An application involving Lorentz-invariance violation
in Newtonian gravity is discussed in Sec.\ \ref{application}.
Spin weight and spin-weighted spherical harmonics
are reviewed in Appendix \ref{harmonics-appendix}.
Appendix \ref{young-appendix} provides
a brief overview of Young symmetrizers.

\section{Construction}
\label{theory}

\subsection{Notation and conventions}
\label{notation}

This section establishes some basic notation
used throughout this paper.
First,
Latin indices $a,b,c,\ldots$ on tensor components
indicate spatial dimensions in one of
the coordinate systems described below.
Greek letters $\al,\be,\ga,\ldots$ are used
in Sec. \ref{application} to
indicate spacetime indices.

Several different special sets of basis vectors are useful.
In addition to the cartesian basis
$\{\Ex, \Ey, \Ez\} = \{\EX, \EY, \EZ\}$,
we define $J_z$-basis vectors
$\{\Eu, \Ed, \Ez\} = \{\ED, \EU, \EZ\}$, where
\beq
\Eu = \ED = \tfrac{1}{\sqrt2}\big(\Ex + i\Ey\big) \ ,\qquad
\Ed = \EU = \tfrac{1}{\sqrt2}\big(\Ex - i\Ey\big) \ .
\label{eud}
\eeq
The standard spherical-coordinate basis vectors are denoted as
$\{\Er, \Eth, \Eph\} = \{\ER, \ETH, \EPH\}$,
where
\beq
\Er = \N\ , \qquad
\Eth = \cos\th\cos\ph\,\Ex +\cos\th\sin\ph\,\Ey  -\sin\th\,\Ez\  ,\qquad
\Eph = -\sin\ph\,\Ex +\cos\ph\,\Ey \ .
\eeq
Finally, we define a helicity basis
$\{\Er, \Ep, \Em\} = \{\ER, \EM, \EP\}$, where
\beq
\Epm = \EMP = \tfrac{1}{\sqrt2}\big(\Eth \pm i\Eph\big) \ .
\eeq
Note that raising and lowering indices
in the $J_z$ basis exchanges ``up'' and ``down'' labels,
while raising and lowering indices in the helicity basis
exchanges ``plus'' and ``minus'' labels.
All bases are defined to be orthonormal:
$\E_a\cdot \E^b = \de_a^b$.
We denote the direction cosines
between two vectors,
not necessarily from the same basis,
as
$g_{aa'} = \E_a\cdot\E_{a'}$,
$g^{aa'} = \E^a\cdot\E^{a'}$, and
$g_a^{a'} = \E_a\cdot\E^{a'}$.
Note that these are the components of the euclidean metric
$g = g^{aa'} \E_a\otimes \E_{a'}$
relative to row basis $\E_a$ and column basis $\E_{a'}$.
In addition to defining the inner product,
the metric components can be used to
transform tensor components between bases,
including the raising and lowering of indices.

We denote the symmetrized tensor product using $\odot$.
The symmetrized product of two vectors $v$ and $u$
is defined to be $v\odot u = \half(v\otimes u + u\otimes v)$.
The $k$-fold symmetrized product of a vector $v$
will be written as $v^{\odot k}$,
which in index notation reads
$(v^{\odot k})^{a_1\ldots a_k}
= \tfrac{1}{k!} v^{(a_1}\ldots v^{a_k)}
= v^{a_1}\ldots v^{a_k}$.
This is the simple $k$-fold tensor product.
The product of products is written
$(v^{\odot k}\odot u^{\odot l})^{a_1 \ldots a_{k+l}}
= \tfrac{1}{(k+l)!}v^{(a_1}\ldots v^{a_k}u^{a_{k+1}}\ldots u^{a_{k+l})}$.
More generally, the product of a rank-$k$ tensor $T$
and a rank-$l$ tensor $S$ is
$(T\odot S)^{a_1 \ldots a_{k+l}}
= \tfrac{1}{(k+l)!}T^{(a_1 \ldots a_k}S^{a_{k+1}\ldots a_{k+l})}$.
We write the $k$-fold symmetric product of a tensor $T$ as $T^{\odot k}$.
The inner product of two equal-rank tensors $T$ and $S$
is defined as the invariant contraction
$T\cdot S = T^{a_1 \ldots a_k}S_{a_1\ldots a_k}$.
Finally,
a tensor index written with an exponent, such as $a^q$,
indicates $q$ copies of the index $a$.
For example, $T^{x^2y^3z} = T^{xxyyyz}$
is a cartesian-basis component of a rank-6 tensor $T$.

\subsection{Scalar spherical harmonics}
\label{scalarY}

In this section,
we develop a new method for calculating
the scalar spherical harmonics  $\yjm{jm}$
in terms of the components of $\N$.
The derivation favors the $J_z$ basis,
in which the components of $\N$ are given by
\beq
\Nu = \tfrac{1}{\sqrt2}(\Nx + i\Ny)
= \tfrac{1}{\sqrt2}\sin\th e^{+ i\ph}\ , \qquad
\Nd = \tfrac{1}{\sqrt2}(\Nx - i\Ny)
= \tfrac{1}{\sqrt2}\sin\th e^{- i\ph}\ , \qquad
\Nz = \cos\th\ .
\label{Jz_nhat}
\eeq
The result of the calculation that follows is
\beq
\yjm{jm}(\N) = 
(-\sgn m)^m
\sqrt{\tfrac{(2j+1)(j+m)!(j-m)!}{4\pi\, 2^{|m|}}}
\sum_{q|_{jm}}
\frac{\Nu^{q_\u}\Nd^{q_\d}\Nz^{q_z}}{(-2)^{\qud}q_\u!q_\d!q_z!} \ ,
\label{Yn}
\eeq
where $q|_{jm}$ is the 
restriction to all nonnegative powers
$q=\{q_\u,q_\d,q_z\}$ that sum to $j$
and obey $q_\u-q_\d=m$,
and $\qud=\min(q_\u,q_\d)$.
In practice,
this can be accomplished by summing over
$q_z = j-|m|, j-|m|-2, j-|m|-4 \ldots \geq 0$,
with the remaining powers set to
$q_\u = \half(j+ m-q_z)$ and
$q_\d = \half(j- m-q_z)$.
For illustrative purposes,
$\yjm{jm}$ up to $j=4$ are given in Table \ref{ytable}.
Combining Eqs.\ \rf{Jz_nhat} and \rf{Yn},
we can write the spherical harmonics in terms of
cartesian components of $\N$.
It also leads to
\beq
\yjm{jm}(\th,\ph) = 
(-\sgn m)^m
\sqrt{\tfrac{(2j+1)(j+m)!(j-m)!}{4\pi\, 2^{|m|}}}\,
e^{im\ph}
\sum_{q|_{jm}}
\frac{(\sin\th)^{q_\u+q_\d}(\cos\th)^{q_z}}
     {2^{(q_\u+q_\d)/2}(-2)^{\qud}q_\u!q_\d!q_z!} \ ,
\label{Yn2}
\eeq
in terms of the spherical-coordinate angles.

The derivation of Eq.\ \rf{Yn} starts
by taking $s_1=s_2=0$, $j_1=1$, $m_1=\pm 1$,
and $m_2=\pm j_2$ in identity \rf{Yproduct}.
This yields the recursion relation
\beq
\yjm{j(\pm j)}(\N)
= \sqrt{\tfrac{2j+1}{j}}\,
\left\{\begin{array}{c} -\Nu \\ \Nd \end{array}\right\}
\yjm{(j-1)(\pm j \mp 1)}(\N)
\label{rec1}\ ,
\eeq
which relates different harmonics at the upper and lower limits
of $m = \pm j$.
Combining this with $\yjm{00}=1/\sqrt{4\pi}$,
we then find
\beq
\yjm{j(\pm j)}(\N)
= \sqrt{\tfrac{(2j+1)!!}{4\pi j!}}\,
\left\{\begin{array}{c} (-\Nu)^j \\ (\Nd)^j \end{array}\right\} \ .
\label{Yjj}
\eeq
We next use ladder operators to find
the harmonics for other values of $m$.

The ladder operators 
$J_\u = \Eu \cdot\mbf J$
and
$J_\d = \Ed \cdot\mbf J$
can be used to respectively
raise and lower the $J_z$-eigenvalue $m$.
When acting on spin-zero scalars,
the ladder operators can be written as the differential operators
$J_\u = \frac{e^{i\ph}}{\sqrt2}(\prt_\th + i\cot\th\prt_\ph)$
and
$J_\d = \frac{e^{-i\ph}}{\sqrt2}(-\prt_\th + i\cot\th\prt_\ph)$.
Acting on the $J_z$-basis components of $\N$,
the ladder operators shift the components according to 
$J_\u \Nu =  -J_\d \Nd = 0$,
$J_\u \Nd =  -J_\d \Nu = \Nz$,
$J_\u \Nz = -\Nu$, and 
$J_\d \Nz =  \Nd$.
As a result,
repeatedly operating on Eq.\ \rf{Yjj} with $J_\d$ or $J_\u$
introduces other components of $\N$,
leaving the total number of components appearing
in the product unchanged.
The $\yjm{jm}$ harmonics 
are then combinations of terms involving
$\Nu^{q_\u}\, \Nd^{q_\d}\, \Nz^{q_z}$,
with powers $q_\u$, $q_\d$,  and $q_z$ that sum to $j$.
Noting that
$J_z = \Ez\cdot\mbf J = -i\prt_\ph$,
we then have
$J_z\Nu = \Nu$, $J_z\Nd = -\Nd$,
and $J_z\Nz =0$,
which gives
$J_z\, \Nu^{q_\u}\, \Nd^{q_\d}\, \Nz^{q_z}
= (q_\u-q_\d)\, \Nu^{q_\u}\, \Nd^{q_\d}\, \Nz^{q_z}$.
So the powers also obey $q_\u-q_\d = m$.
This implies that $q_z$ is restricted to
$j-|m|, j-|m|-2, j-|m|-4 \ldots \geq 0$,
with the remaining powers given by
$q_\u = \half(j + m-q_z)$ and
$q_\d = \half(j - m-q_z)$.
The harmonics then take the form
\beq
\yjm{jm}(\N)
= \sum_{q_z} \Aqjm{q_z}{jm}\,
\Nu^{\frac12(j+m-q_z)}\, \Nd^{\frac12(j-m-q_z)}\, \Nz^{q_z} \ ,
\eeq
where the constant coefficients $\Aqjm{q_z}{jm}$
are nonzero for the values of $q_z$ given above.

Equation \rf{Yjj} implies that the nonzero
coefficients for $m=\pm j$ are
$\Aqjm{0}{j(\pm j)}
= (\mp 1)^j\sqrt{(2j+1)!!/4\pi j!}$\,.
To find the other $\Aqjm{q_z}{jm}$ coefficients,
we adopt the conventional normalization 
\beq
\left\{\begin{array}{c}J_\u\\ J_\d\end{array}\right\}
\yjm{jm} = \sqrt{\tfrac12(j\pm m+1)(j\mp m)}\, \yjm{j(m\pm 1)} \ .
\eeq
Ladder operations then lead to the recursion relation
\beq
\sqrt{\tfrac12(j\pm m+1)(j\mp m)}\, \Aqjm{q_z}{j(m\pm1)}
= \pm\half(j\mp m-q_z+1) \Aqjm{q_z-1}{jm}
\mp (q_z+1) \Aqjm{q_z+1}{jm} \ .
\eeq
By replacing $m\rightarrow \mp (|m|+1)$
and taking $q_z = j-|m|$,
we get a recursion for
cases in which either $q_\u$ or $q_\d$ vanish,
which leads to
\beq
\Aqjm{j-|m|}{jm}
=\tfrac{(-\sgn m)^m}{|m|!}
\sqrt{\tfrac{(2j+1)}{4\pi\, 2^{|m|}}\tfrac{(j+|m|)!}{(j-|m|)!}} \ .
\label{Cqmax}
\eeq
We find a closed-form expression
for the remaining coefficients
by combining the raising and lowering relations
to get a recursion between coefficients with the same $m$:
\beq
0 =
(q_z+1)(q_z+2)\,\Aqjm{q_z+2}{jm}
+\tfrac{(j+m-q_z)(j-m-q_z)+(q_z-1)q_z}{2}\,\Aqjm{q_z}{jm}
+\tfrac{(j+m-q_z+2)(j-m-q_z+2)}{4}\,\Aqjm{q_z-2}{jm} \ .
\eeq
Defining
\beq
\Bqjm{q_z}{jm}
= (q_z+1)(q_z+2)\,\Aqjm{q_z+2}{jm}
+ \tfrac{(j+m-q_z)(j-m-q_z)}{2}\,\Aqjm{q_z}{jm} \ ,
\eeq
the recursion relation can be written as
$\Bqjm{q_z-2}{jm} = -2\Bqjm{q_z}{jm}$.
This implies that all the $\Bqjm{q_z}{jm}$ constants
are proportional to $\Bqjm{j-|m|}{jm}$, which is zero.
Therefore, all $\Bqjm{q_z}{jm}$ constants vanish,
and we have
\beq
\Aqjm{q_z}{jm}
= -\tfrac{2(q_z+1)(q_z+2)}{(j+m-q_z)(j-m-q_z)}
\Aqjm{q_z+2}{jm} \ .
\eeq
With this we can write all of the coefficients in
terms of those given in Eq.\ \rf{Cqmax}.
The result reduces to
\beq
\Aqjm{q_z}{jm}
= \tfrac{(-\sgn m)^m}{(-2)^{\qud}q_\u!q_\d!q_z!}
\sqrt{\tfrac{(2j+1)(j+m)!(j-m)!}{4\pi\, 2^{|m|}}} \ ,
\eeq
where $q_\u = \half(j+ m-q_z)$,
$q_\d = \half(j- m-q_z)$,
and $\qud=\min(q_\u,q_\d)$.
We then arrive at Eq.\ \rf{Yn}.

\begin{table}
  \renewcommand{\tabcolsep}{5pt}
  \renewcommand{\arraystretch}{1.8}
  \begin{tabular}{c|c|c}
    $jm$ &
    $\yjm{jm}(\N)$ &
    $\Yjm{jm}$ \\
    \hline\hline
    00 &
    $\sqrt{\tfrac{1}{4\pi}}$ &
    $1$ \\
    \hline
    10 &
    $\sqrt{\tfrac{3}{4\pi}}\Nz$ &
    $\Ez$ \\
    11 &
    $-\sqrt{\tfrac{3}{4\pi}}\Nu$ &
    $-\Eu$ \\
    \hline
    20 &
    $\sqrt{\tfrac{5}{4\pi}}(\Nz^2-\Nu\Nd)$ &
    $\sqrt{\tfrac23}(\Ez\odot\Ez -\Eu\odot\Ed)$ \\
    21 &
    $-\sqrt{\tfrac{15}{4\pi}}\Nu\Nz$ &
    $-\sqrt2\, \Eu\odot\Ez$ \\
    22 & $\sqrt{\tfrac{15}{8\pi}}\Nu^2$ &
    $\Eu\odot\Eu$ \\
    \hline
    30 &
    $\sqrt{\tfrac{7}{4\pi}}\big(\Nz^3-3\Nu\Nd\Nz\big)$ &
    $\sqrt{\tfrac{2}{5}}\big(\Ez\odot\Ez\odot\Ez
    -3\Eu\odot\Ed\odot\Ez\big)$\\
    31 &
    $-\sqrt{\tfrac{21}{8\pi}}\big(2\Nu\Nz^2-\Nu^2\Nd\big)$ &
    $-\sqrt{\tfrac{3}{5}}\big(2\Eu\odot\Ez\odot\Ez
    -\Eu\odot\Eu\odot\Ed\big)$ \\
    32 &
    $\sqrt{\tfrac{105}{8\pi}}\Nu^2\Nz$ &
    $\sqrt3\, \Eu\odot\Eu\odot\Ez$\\
    33 &
    $-\sqrt{\tfrac{35}{8\pi}}\Nu^3$ &
    $-\Eu\odot\Eu\odot\Eu$\\
    \hline
    40 & $\sqrt{\tfrac{9}{16\pi}}(2\Nz^4-12\Nu\Nd\Nz^2+3\Nu^2\Nd^2)$ &
    $\sqrt{\tfrac{2}{35}}(2\Ez\odot\Ez\odot\Ez\odot\Ez
    -12 \Eu\odot\Ed\odot\Ez\odot\Ez
    +3 \Eu\odot\Eu\odot\Ed\odot\Ed)$\\
    41 & $-\sqrt{\tfrac{45}{8\pi}}(2\Nu\Nz^3-3\Nu^2\Nd\Nz)$ &
    $-\sqrt{\tfrac{4}{7}}(
    2\Eu\odot\Ez\odot\Ez\odot\Ez
    -3\Eu\odot\Eu\odot\Ed\odot\Ez)$\\
    42 & $\sqrt{\tfrac{45}{8\pi}}(3\Nu^2\Nz^2-\Nu^3\Nd)$ &
    $\sqrt{\tfrac{4}{7}}(
    3\Eu\odot\Eu\odot\Ez\odot\Ez
    -\Eu\odot\Eu\odot\Eu\odot\Ed)$\\
    43 & $-\sqrt{\tfrac{315}{8\pi}}\Nu^3\Nz$ &
    $-2\Eu\odot\Eu\odot\Eu\odot\Ez$\\
    44 & $\sqrt{\tfrac{315}{32\pi}}\Nu^4$ &
    $\Eu\odot\Eu\odot\Eu\odot\Eu$\\
  \end{tabular}
  \caption{Spherical harmonics and
    the  traceless spherical-harmonic tensors for $j\leq 4$.
    Only the nonnegative $m$ cases are shown.
    The $\yjm{jm}$ for negative $m$ can be found using
    $\yjm{j(-m)} = (-1)^m \yjm{jm}^*$,
    which results in the replacement
    $\protect\Nu\leftrightarrow\protect\Nd$
    and multiplication by the Condon-Shortley phase $(-1)^m$.
    The $\Yjm{jm}$ for negative $m$ can be found using
    $\Yjm{j(-m)} = (-1)^m \Yjm{jm}^*$,
    resulting in the replacement
    $\protect\Eu\leftrightarrow\protect\Ed$
    and multiplication by $(-1)^m$.
    \label{ytable}}
\end{table}

\subsection{Traceless spherical-harmonic tensors}
\label{Ytensors}

Next, we extract
the orthonormal rank-$j$ symmetric traceless tensors $\Yjm{jm}$
and discuss their properties.
Notice that Eq.\ \rf{Yn} can be written as
the inner product of two rank-$j$ tensors,
\beq
\yjm{jm}(\N) = 
\sqrt{\tfrac{(2j+1)!!}{4\pi\, j!}}\
\Yjm{jm}\cdot \N^{\odot j} \ ,
\label{Y0}
\eeq
where the spherical-harmonic tensors are defined as
\beq
\Yjm{jm} = 
(-\sgn m)^m
\sqrt{\tfrac{j!(j+m)!(j-m)!}{2^{|m|}(2j-1)!!}}
\sum_{q|_{jm}}
\frac{
  \Eu^{\odot q_\u} \odot
  \Ed^{\odot q_\d} \odot
  \Ez^{\odot q_z}
}{(-2)^{\qud}q_\u!q_\d!q_z!} \ .
\label{Yjm}
\eeq
Examples of spherical-harmonic tensors
for $j\leq4$ are included in Table \ref{ytable}.
While these are conveniently expressed in
terms of the $J_z$-basis vectors,
they can be written in
the cartesian basis using Eq.\ \rf{eud}.
More generally,
the components of $\Yjm{jm}$ in any basis $\E_a$
can be written in terms of the direction cosines between
the $\E_a$ vectors and the $J_z$-basis vectors:
\bea
(\Yjm{jm})^{a_1a_2\ldots a_j} &=&
\Yjm{jm} \cdot (\E^{a_1}\otimes \E^{a_2}\otimes \ldots \otimes \E^{a_j})
\notag \\ &=&
(-\sgn m)^m
\sqrt{\tfrac{(j+m)!(j-m)!}{2^{|m|}j!(2j-1)!!}}
\sum_{q|_{jm}}
\frac{
  g_\u^{(a_1}\ldots g_\u^{a_{q_\u}}
  g_\d^{a_{q_\u+1}}\ldots g_\d^{a_{q_\u+q_\d}}
  g_z^{a_{q_\u+q_\d+1}}\ldots g_z^{a_j)}
}{(-2)^{\qud}q_\u!q_\d!q_z!} \ .
\label{Yjmcomps}
\eea
The complex conjugate of $\Yjm{jm}$ is given by
\beq
\Yjm{jm}^* = (-\sgn m)^m \sqrt{\tfrac{j!(j+m)!(j-m)!}{2^{|m|}(2j-1)!!}}
\sum_{q|_{jm}}
\frac{
  \EU{}^{\odot q_\u} \odot
  \ED{}^{\odot q_\d} \odot
  \EZ{}^{\odot q_z}
}{(-2)^{\qud}q_\u!q_\d!q_z!} \ .
\label{Ycc}
\eeq
Note that the $\Yjm{jm}$ include the Condon-Shortley phase
and obey the relation
\beq
\Yjm{jm}^* = (-1)^m\Yjm{j(-m)} \ .
\eeq
Below we show that
the spherical-harmonic tensors obey the orthonormality relation
\beq
\Yjm{jm}\cdot\Yjm{jm'}^* = \de_{mm'}
\label{Yorth}
\eeq
and serve as an orthonormal basis
for the $(2j+1)$-dimensional
space of rank-$j$ symmetric traceless tensors
in three dimensions.

Using the $\Yjm{jm}$,
we can perform a spherical decomposition
of an arbitrary symmetric traceless rank-$j$ tensor $T$,
\beq
T = \sum_{m} T_{jm} \Yjm{jm} \ .
\eeq
The spherical-expansion coefficients are given by
the inner product with the conjugate basis tensor,
\beq
T_{jm} = \Yjm{jm}^*\cdot T
= (-\sgn m)^m \sqrt{\tfrac{j!(j+m)!(j-m)!}{2^{|m|}(2j-1)!!}}
\sum_{q|_{jm}}
\frac{T^{\u^{q_\u} \d^{q_\d} z^{q_z}}
}{(-2)^{\qud}q_\u!q_\d!q_z!} \ .
\eeq
The $T$ tensor and the basis tensors $\Yjm{jm}$ are spin-$j$ objects.
Each component has the same $j$ value
but can have different $m$ values.
The $T_{jm}$ give the components
with fixed $J_z=m$.

The remainder of this section
is devoted to proving that the $\Yjm{jm}$
are traceless and orthonormal.
To show that they are traceless,
we first note that the trace of
$\Eu^{\odot q_\u} \odot
\Ed^{\odot q_\d} \odot
\Ez^{\odot q_z}$
is
\beq
\tfrac{q_z(q_z-1)}{j(j-1)}
\Eu^{\odot q_\u}\odot
\Ed^{\odot q_\d}\odot
\Ez^{\odot(q_z-2)}
+
\tfrac{2q_\u q_\d}{j(j-1)}
\Eu^{\odot(q_\u-1)}\odot
\Ed^{\odot(q_\d-1)}\odot
\Ez^{\odot q_z}\ .
\eeq
So the trace of Eq.\ \rf{Yjm} is proportional to
\beq
\sum_{q|_{jm}}
\frac{
\Eu^{\odot q_\u}\odot
\Ed^{\odot q_\d}\odot
\Ez^{\odot(q_z-2)}
}{(-2)^{\qud}q_\u!q_\d!(q_z-2)!}
+ 2\sum_{q|_{jm}}
\frac{\Eu^{\odot(q_\u-1)}\odot
\Ed^{\odot(q_\d-1)}\odot\Ez^{\odot q_z}}
{(-2)^{\qud}(q_\u-1)!(q_\d-1)!q_z!}
=  0 \ ,
\eeq
proving that the $\Yjm{jm}$ tensors
are traceless.

To show orthonormality,
we contract $\Yjm{jm}$ with $\Yjm{jm'}^*$.
The orthogonality of the $\{\Eu,\Ed,\Ez\}$ basis
implies that the only nonzero terms in the resulting double
sum are those with matching $q$ powers.
This immediately implies
$\Yjm{jm}$ tensors with different $m$ values
are orthogonal.
A short calculation then shows that
the inner product of two $\Yjm{jm}$ tensors reduces to
\beq
\Yjm{jm}\cdot\Yjm{jm'}^*
= \de_{mm'}
\tfrac{(j+m)!(j-m)!}{2^{|m|}(2j-1)!!}
\sum_{q|_{jm}}
\frac{1}{4^{\qud}q_\u!q_\d!q_z!} \ .
\label{YY}
\eeq
It is then useful to relabel
$q_1=\qud=\min(q_\u,q_\d)$,
$q_2=\max(q_\u,q_\d)$,
and
$q_3=q_z$.
The sum in the above expression can be written as
$\sum_{q|_{jm}} \frac{1}{q_1!q_2!q_3!}\big(\tfrac14\big)^{q_1}$
and is restricted to
$q_1+q_2+q_3=j$ and $q_2-q_1=|m|$.
It can be evaluated by considering
the multinomial expansion
\beq
\tfrac{1}{j!}\big(\tfrac1{4z} + z + 1\big)^j
= \sum_{q|_j}
\tfrac{1}{q_1!q_2!q_3!} (\tfrac{1}{4})^{q_1} z^{q_2-q_1}
= \sum_{m=-j}^j \Cm{m} z^m\ ,
\eeq
where $q|_j$ is the restriction to sets
of nonnegative powers $\{q_1,q_2,q_3\}$ adding to $j$.
The $\Cm{m}$ expansion coefficients are
partial sums further restricted by $q_2-q_1 = m$.
This implies that the sum in Eq.\ \rf{YY}
is equivalent to the coefficient $C_{|m|}$.
We can calculate the $\Cm{m}$ coefficients using the contour integral
\beq
\Cm{m} = \tfrac{1}{2\pi i}\oint_\ga \tfrac{1}{j!}\big(\tfrac1{4z} + z + 1\big)^j z^{-(m+1)}\, dz
=\tfrac{2^m (2j-1)!!}{(j+m)!(j-m)!} \ ,
\eeq
where $\ga$ is any counterclockwise contour enclosing
the origin in the complex plane.
Along with Eq.\ \rf{YY},
this result implies the $\Yjm{jm}$ tensors are orthonormal.

\subsection{Generalized spherical-harmonic tensors}
\label{Yrtensors}

By taking symmetric products
of the $\Yjm{jm}$ tensors with the metric $g$,
we can generalize the $\Yjm{jm}$
to create a basis for symmetric
rank-$\r$ tensors including traces.
Each metric increases the rank by two,
so the number of metric tensors in
the product is $\frac12(\r-j)$.
We then define
\bea
\Yrjm{\r}{jm}
&=& \sqrt{\tfrac{\r!(2j+1)!!}{j!(\r+j+1)!!(\r-j)!!}}\
\Yjm{jm} \odot g^{\odot \frac12(\r-j)} 
\notag \\
&=&
(-\sgn m)^m
\sqrt{\tfrac{(2j+1)\r!(j+m)!(j-m)!}{2^{|m|}(\r+j+1)!!(\r-j)!!}}
\sum_{q|_{jm}}
\frac{
  \Eu^{\odot q_\u} \odot
  \Ed^{\odot q_\d} \odot
  \Ez^{\odot q_z} \odot
  g^{\odot \frac12(\r-j)}
}{(-2)^{\qud}q_\u!q_\d!q_z!} \ .
\label{Yrjm}
\eea
These form an orthonormal basis
for symmetric rank-$\r$ tensors including trace elements.
The $jm$ indices give the spin angular momentum
of the basis tensor.
The components in any basis $\E_a$ are given by
\bea
(\Yrjm{\r}{jm})^{a_1a_2\ldots a_\r}
&=&
(-\sgn m)^m
\sqrt{\tfrac{(2j+1)(j+m)!(j-m)!}{2^{|m|}\r!(\r+j+1)!!(\r-j)!!}}
\notag \\
&&\quad\times
\sum_{q|_{jm}}
\frac{
  g_\u^{(a_1}\ldots g_\u^{a_{q_\u}}
  g_\d^{a_{q_\u+1}}\ldots g_\d^{a_{q_\u+q_\d}}
  g_z^{a_{q_\u+q_\d+1}}\ldots g_z^{a_j}
  g_{\phantom{z}}^{a_{j+1}a_{j+2}} \ldots g_{\phantom{z}}^{a_{\r-1}a_{\r})}
}{(-2)^{\qud}q_\u!q_\d!q_z!} \ .
\label{Yrjmcomps}
\eea
In terms of the rank-$\r$ spherical-harmonic tensors,
the scalar spherical harmonics are
\beq
\yjm{jm}(\N) = 
\sqrt{\tfrac{(\r+j+1)!!(\r-j)!!}{4\pi\, \r!}}\
\Yrjm{\r}{jm}\cdot \N^{\odot \r} \ ,
\label{Y0r}
\eeq
providing a generalization of Eq.\ \rf{Y0}.
The conjugate tensors are
\beq
\Yrjmconj{\r}{jm}
= (-\sgn m)^m
\sqrt{\tfrac{(2j+1)\r!(j+m)!(j-m)!}{2^{|m|}(\r+j+1)!!(\r-j)!!}}
\sum_{q|_{jm}}
\frac{
  \EU{}^{\odot q_\u} \odot
  \ED{}^{\odot q_\d} \odot
  \EZ{}^{\odot q_z} \odot
  g^{\odot \frac12(\r-j)} 
}{(-2)^{\qud}q_\u!q_\d!q_z!}
\label{Yrjmconj}
\eeq
and satisfy
\beq
\Yrjmconj{\r}{jm} = (-1)^m \Yrjm{\r}{j(-m)} \ .
\label{Yrcc}
\eeq
The $\Yrjm{\r}{jm}$ of the same rank are orthonormal,
\beq
\Yrjm{\r}{jm}\cdot\Yrjmconj{\r}{j'm'} = \de_{jj'}\de_{mm'} \ ,
\label{Yrorth}
\eeq
which we prove below.
Note that for fixed $\r$,
the values of $j$ are restricted to $j=\r,\r-2,\r-4,\ldots \geq 0$,
and the total number of $\Yrjm{\r}{jm}$ tensors is $\sum (2j+1) = \frac12(\r+1)(\r+2)$,
matching the dimension of the space
of symmetric rank-$\r$ tensors.
We recover
the traceless spherical-harmonic tensors
$\Yjm{jm}=\Yrjm{j}{jm}$
when $j=\r$.
For even $\r$, $j=m=0$
corresponds to the normalized total-trace element
$\Yrjm{\r}{00} = g\odot\ldots\odot g /\sqrt{\r+1}$.

Any symmetric rank-$\r$ tensor $T$ can be expanded
in the generalized spherical-harmonic tensors,
\beq
T = \sum_{jm} T_{jm} \Yrjm{\r}{jm} \ ,
\eeq
where the spherical components are
\beq
T_{jm}
= \Yrjmconj{\r}{jm}\cdot T
= (-\sgn m)^m
\sqrt{\tfrac{(2j+1)\r!(j+m)!(j-m)!}{2^{|m|}(\r+j+1)!!(\r-j)!!}}
\sum_{q|_{jm}}
\frac{{T^{\u^{q_\u}\d^{q_\d} z^{q_z}a_1\ldots a_{(\r-j)/2}}}_{a_1\ldots a_{(\r-j)/2}}
}{(-2)^{\qud}q_\u!q_\d!q_z!} \ .
\label{Tcomps}
\eeq
This provides both an angular-momentum decomposition
and a trace decomposition.
The trace decomposition can be written
\beq
T = \sum_j \sqrt{\tfrac{\r!(2j+1)!!}{j!(\r+j+1)!!(\r-j)!!}}\
T_j \odot g^{\odot\frac12(\r-j)} \ ,
\eeq
where
\beq
T_j = \sum_m T_{jm} \Yjm{jm} 
\eeq
are rank-$j$, symmetric, and traceless.

Definition \rf{Yrjm} implies that
the generalized spherical-harmonic tensors
of different rank are related through the recursion relation
\beq
\Yrjm{\r}{jm}
= \sqrt{\tfrac{\r(\r-1)}{(\r+j+1)(\r-j)}}\ g\odot \Yrjm{\r-2}{jm} \ .
\label{YrjmAntiTrace}
\eeq
They are also connected by the trace identity
\beq
g\cdot\Yrjm{\r}{jm} = \sqrt{\tfrac{(\r+j+1)(\r-j)}{\r(\r-1)}}\ \Yrjm{\r-2}{jm}\ ,
\label{YrjmTrace}
\eeq
where the dot $\cdot$ indicates
the contraction of $g$
with any two indices of $\Yrjm{\r}{jm}$,
yielding a tensor of rank $\r-2$.

We prove trace relation \rf{YrjmTrace}
by first considering
the contraction of the metric $g$
with the symmetric product $g\odot T$,
where $T$ is a symmetric rank-$\r$ tensor.
A calculation yields the identity
\beq
g\cdot(g\odot T) = \Dr{\r}\, T + \Fr{\r}\, g\odot (g\cdot T) \ ,
\eeq
where
$\Dr{\r} = 2(2\r+3)/(\r+2)(\r+1)$ and
$\Fr{\r} = \r(\r-1)/(\r+2)(\r+1)$.
For fixed $j$ and $m$,
we apply this to the tensors
\beq
\Tr{\r}
= g\odot \Tr{\r-2}
= g^{\odot\frac12(\r-j)}\odot \Yjm{jm}
= \sqrt{\tfrac{j!(\r+j+1)!!(\r-j)!!}{\r!(2j+1)!!}}\ \Yrjm{\r}{jm} \ ,
\eeq
which gives
\beq
g\cdot \Tr{\r} = g\cdot(g\odot \Tr{\r-2})
= \Dr{\r-2}\, \Tr{\r-2} +\Fr{\r-2}\, g\odot (g\cdot \Tr{\r-2})\ .
\eeq
Iterating, we find
\bea
g\cdot \Tr{\r}
&=&
(\Dr{\r-2}
+\Fr{\r-2}\Dr{\r-4}
+\Fr{\r-2}\Fr{\r-4}\Dr{\r-6}
+ \ldots 
+\Fr{\r-2}\ldots\Fr{j+2}\Dr{j} )\ \Tr{\r-2}
\notag \\
&=&
\tfrac{(\r+j+1)(\r-j)}{\r(\r-1)}\Tr{\r-2} \ ,
\eea
which leads to Eq.\ \rf{YrjmTrace}.

Next, identities \rf{YrjmAntiTrace} and \rf{YrjmTrace}
can be used to show that the $\Yrjm{\r}{jm}$ tensors are orthonormal.
We first note that orthogonality
follows immediately from
the tracelessness and orthogonality of the $\Yjm{jm}$.
The inner product $\Yrjm{\r}{jm}\cdot\Yrjmconj{\r}{j'm'}$
involves traces of $\Yjm{jm}$ unless $j=j'$
and is proportional to $\Yjm{jm}\cdot\Yjm{jm'}^* = \de_{mm'}$
when $j=j'$.
Then using Eqs.\ \rf{YrjmAntiTrace} and \rf{YrjmTrace},
we can write the inner product as
\bea
\Yrjm{\r}{jm}\cdot\Yrjmconj{\r}{jm}
&=&
\sqrt{\tfrac{\r(\r-1)}{(\r+j+1)(\r-j)}}\
\big(g\odot \Yrjm{(\r-2)}{jm}\big)
\cdot\Yrjmconj{\r}{jm}
\notag \\
&=&
\sqrt{\tfrac{\r(\r-1)}{(\r+j+1)(\r-j)}}\
\Yrjm{(\r-2)}{jm}
\cdot(g\cdot\Yrjmconj{\r}{jm})
\notag \\
&=&
\Yrjm{(\r-2)}{jm} \cdot\Yrjmconj{\r-2}{jm} \ ,
\eea
which implies all $\Yrjm{\r}{jm}$ are normalized
since the lowest-rank $\Yrjm{j}{jm}=\Yjm{jm}$ are normalized.

\subsection{Spin-weighted spherical harmonics}
\label{tensorY}

The goal of this section is to show
that each helicity-basis component of
the spherical-harmonic tensor $\Yrjm{\r}{jm}$
is proportional to a spin-weighted
spherical harmonic $\syjm{s}{jm}$.
See Appendix \ref{harmonics-appendix}
for a review of spin weight and spin-weighted harmonics.
The key result of the calculation that follows is 
\beq
(\Yrjm{\r}{jm})_{+^{q_+} -^{q_-} r^{q_{r\phantom{+}}} }
= \Yrjm{\r}{jm}
\cdot(\Ep^{\odot q_+}\odot\Em^{\odot q_-}\odot\Er^{\odot q_r})
= \sNrqj{s}{\r}{\qpm}{j}\ \syjm{s}{jm} \ ,
\label{Ysr}
\eeq
where it is understood that $\r=q_++q_-+q_r$,
$s=q_+-q_-$, and $\qpm = \min(q_+,q_-)$.
In practice,
the constraints can be handled by taking
$q_r =\r-|s|, \r-|s|-2, \r-|s|-4,\ldots \geq0$,
with the remaining powers set to 
$q_\pm = \half(\r\pm s-q_r)$.
The normalization constants $\sNrqj{s}{\r}{\qpm}{j}$
are zero unless $\r-j = \text{even} \geq 0$,
and the nonzero values are given by
\bea
\sNrqj{s}{\r}{\qpm}{j} &=&
(-\sgn s)^s
\frac{(\r-|s|)!}{(-2)^{\qpm}}
\sqrt{\frac{4\pi(j+|s|)!}{2^{|s|}(j-|s|)!\r!(\r+j+1)!!(\r-j)!!}}
\notag\\
&&\quad
\times\
{}_3F_2\big(-\qpm,-\tfrac{\r-j}{2},-\tfrac{\r+j+1}{2};
-\tfrac{\r-|s|}{2},-\tfrac{\r-|s|-1}{2};1\big)\ ,
\label{Nq}
\eea
where ${}_3F_2$ is a generalized hypergeometric function.
The special case in Eq.\ \rf{Y0r} corresponds to $q_+=q_-=0$,
and Eq.\ \rf{Y0} is further restricted to $\r=j$.
Note that some powers $\{q_+,q_-,q_r\}$ obeying
the above restrictions give vanishing
$\sNrqj{s}{\r}{\qpm}{j}$ constants,
implying that the corresponding components of $\Yrjm{\r}{jm}$ are zero.

Equation \rf{Ysr} provides an alternative method for calculating
the spin-weighted spherical harmonics.
For example,
we can relate the spin-weighted harmonics $\syjm{s}{jm}$ to the 
trace-free rank-$j$ spherical-harmonic tensors $\Yjm{jm}$
by setting $\r=j$.
We can also take $q_+=0$ or $q_-=0$,
which leads to the comparatively simple relation
\beq
\syjm{s}{jm} = (-\sgn s)^s
\sqrt{\tfrac{\smash{2^{|s|}}j!(2j+1)!!}{4\pi (j+s)!(j-s)!}}\,
\Yjm{jm}\cdot (\Epm^{\odot|s|}\odot\Er^{\odot(j-|s|)}) \ ,
\eeq
where the sign on the $\Epm$ basis vector is the sign of $s$.
This generalizes Eq.\ \rf{Y0} to nonzero spin weight.

The derivation of Eqs.\ \rf{Ysr} and \rf{Nq}
starts by using spin-weight ladder operators
$J_\pm = \Epm\cdot\mbf J
= \tfrac{1}{\sqrt2}(\pm\prt_\th +i \csc\th\,\prt_\ph - s \cot\th)$
to raise and lower the spin weight of Eq.\ \rf{Y0r}.
With the conventional normalization,
the spin-weighted harmonics are related through \cite{km09}
\beq
J_\pm\ \syjm{s}{jm} = -\sqrt{\tfrac12(j\pm s+1)(j\mp s)}\ \syjm{s\pm1}{jm}\ .
\eeq
Applying $J_\pm$ to the right-hand side of Eq.\ \rf{Y0r},
we find that $J_\pm$ converts one $\N=\Er$ to $\Epm$.
This raises $q_\pm$ by one
and lowers $q_r$ by one,
incrementing the spin weight $s=q_+-q_-$
but leaving $\r=q_++q_-+q_r$ unchanged.
Repeated application of ladder operators
produces an expression of the form of Eq.\ \rf{Ysr}
for the special case in which either $q_+$ or $q_-$ is zero,
yielding
\beq
\sNrqj{s}{\r}{0}{j}
= (-\sgn s)^{s} (\r-|s|)!
\sqrt{\tfrac{4\pi (j+|s|)!}{2^{|s|}\r! (j-|s|)! (\r+j+1)!!(\r-j)!!}} \ .
\label{Nq0}
\eeq

For cases where $q_+$ and $q_-$ are both nonzero,
we use the completeness relation
$g = 2 \Ep\odot\Em +\Er\odot\Er$
to write Eq.\ \rf{Ysr} as
\bea
\sNrqj{s}{\r}{\qpm}{j}\syjm{s}{jm}&=&
\Yrjm{\r}{jm}\cdot
(\Epm^{\odot|s|} \odot
\Er^{\odot\r-|s|-2\qpm}\odot
\Ep^{\odot \qpm} \odot
\Em^{\odot \qpm} )
\notag \\
&=&
\big(\half\big)^{\qpm}\,
\Yrjm{\r}{jm}\cdot
\big(\Epm^{\odot|s|} \odot
\Er^{\odot\r-|s|-2\qpm}\odot
  (g-\Er\odot\Er)^{\odot \qpm}
\big)
\notag \\
&=&
\big(-\half\big)^{\qpm}
\sum_l\bc{\qpm}{l}(-1)^l\,
\Yrjm{\r}{jm}\cdot
\big(\Epm^{\odot|s|} \odot
\Er^{\odot\r-|s|-2l}\odot
g^{\odot l}
\big) \ ,
\eea
where $\bc{p}{q}$ are binomial coefficients,
and the index on the $\Epm$ basis vector
matches the sign of $s$.
The sum is limited to
$0\leq l\leq \min\big(\qpm,\half(\r-j)\big)$,
where the limit $\half(\r-j)$
is due to the tracelessness of $\Yrjm{j}{jm}$.
Using trace identity \rf{YrjmTrace}
and Eq.\ \rf{Ysr}, we can write
\bea
\sNrqj{s}{\r}{\qpm}{j}\syjm{s}{jm}&=&
\big(-\half\big)^{\qpm}
\sum_l\bc{\qpm}{l}
(-1)^l
\sqrt{\tfrac{(\r-2l)!(\r+j+1)!!(\r-j)!!}{\r!(\r+j+1-2l)!!(\r-j-2l)!!}}
\Yrjm{\r-2l}{jm}\cdot
\big(
\Epm^{\odot|s|} \odot
\Er^{\odot\r-|s|-2l}
\big)
\notag \\
&=&
\big(-\half\big)^{\qpm}
\sum_l\bc{\qpm}{l}
(-1)^l
\sqrt{\tfrac{(\r-2l)!(\r+j+1)!!(\r-j)!!}{\r!(\r+j+1-2l)!!(\r-j-2l)!!}}
\sNrqj{s}{(\r-2l)}{0}{j}
\syjm{s}{jm} \ ,
\eea
which leads to
\beq
\sNrqj{s}{\r}{\qpm}{j} =
(-\sgn s)^{s}
\big(-\half\big)^{\qpm}
\sqrt{\tfrac{4\pi (j+|s|)!(\r+j+1)!!(\r-j)!!}{2^{|s|} \r! (j-|s|)!}}
\sum_l\bc{\qpm}{l}
\tfrac{(-1)^l(\r-|s|-2l)!}{(\r+j+1-2l)!!(\r-j-2l)!!}\ .
\label{N}
\eeq
Manipulating the factorials,
one can show that the sum in
this expression is equivalent to
\beq
\tfrac{(\r-|s|)!}{(\r-j)!!(\r+j+1)!!}
{}_3F_2\big(-\qpm,-\tfrac{\r-j}{2},-\tfrac{\r+j+1}{2};
-\tfrac{\r-|s|}{2},-\tfrac{\r-|s|-1}{2};1\big)\ ,
\eeq
which implies Eq.\ \rf{Nq}.

\section{Illustrative Examples}
\label{examples}

The formalism developed in this work
can be used to perform a full trace and angular-momentum
decomposition of any tensor or tensor-valued
function in three dimensions.
As a simple example,
consider the scalar function
of the position vector $\mbf r = x\Ex + y\Ey + z\Ez = r \N$
given by
\beq
f(\mbf r)
= x^2 + 2yz
= r^2\big( \half \n_\u^2 + \half \n_\d^2 +\n_\u \n_\d
- i \sqrt2\, \n_\u \n_z + i \sqrt2\, \n_\d \n_z\big) \ .
\eeq
We can write this as
\beq
f(\mbf r)
= r^2 T^{ab} \n_a \n_b \ ,
\eeq
where the tensor $T$ can be taken as symmetric with
nonzero cartesian components $T^{xx} = T^{yz} =T^{zy} = 1$.
In the $J_z$ basis,
the nonzero components are
$T^{\u\u} = T^{\d\d} = T^{\u\d} = \half$
and
$T^{\d z} = -T^{\u z} = i/\sqrt2$.
The spherical-harmonic expansion
$f = \sum_{jm}f_{jm}\yjm{jm}(\N)$
can be found by first expanding $T$ in the basis
of rank-2 spherical-harmonic tensors:
$T = \sum_{jm} T_{jm} \Yrjm{2}{jm}$.
The spherical components are the projections
$T_{jm} = \Yrjmconj{2}{jm}\cdot T$,
which can be calculated using Eq.\ \rf{Tcomps}.
The result is
\beq
T_{2(\pm 2)} = \half\ , \quad
T_{2(\pm 1)} = i\ , \quad
T_{20} = -\tfrac1{\sqrt6} \ , \quad
T_{00} = \tfrac1{\sqrt3}\ .
\eeq
Note that the $j=2$ components give the traceless part of $T$,
while $j=0$ is the trace component.
Using Eq.\ \rf{Y0r} or Eq.\ \rf{Ysr},
we can construct the spherical-harmonic expansion of $f$:
\beq
f(\mbf r)
= \sum_{jm} r^2 T_{jm} \Yrjm{2}{jm}\cdot \N^{\odot 2}
= \sum_{jm} r^2 T_{jm}\, \sNrqj{0}{2}{0}{j}\, \yjm{jm}(\N)\ .
\eeq
So the spherical-harmonic coefficients for the function $f$ are
\beq
f_{jm} = r^2 T_{jm}\, \sNrqj{0}{2}{0}{j}
= r^2 \sqrt{\tfrac{8\pi}{(j+3)!!(2-j)!!}}\, T_{jm}\ .
\eeq
Note that these techniques allow for
the algebraic construction of
the spherical-harmonic expansion,
providing an alternative to
the standard method,
where the coefficients are calculated through
the solid-angle integrals
$f_{jm} = \int \yjm{jm}^* f\, \sin\th\, d\th\, d\ph$.

The scalar $f$ and the tensor $T$ in the above example
both contain components with total angular momentum
$j=0$ and $j=2$.
While $f$ has orbital angular momentum
and $T$ has spin angular momentum,
there is a third object involving $T$
that incorporates both spin and orbital angular momentum.
The vector $\mbf V(\mbf r) = T\cdot\mbf r$
is a spin-1 function with orbital angular momentum.
Its helicity-basis components are spin-weighted functions
and can be expanded in spin-weighted spherical harmonics.
The radial component
$V_r = \Er\cdot\mbf V$
can be expanded in the usual $s=0$ spherical harmonics,
while the $V_\pm  = \Epm\cdot\mbf V$ components
are expanded in $s=\pm1$ harmonics.
The result is
\bea
V_r &=& r \sum_{jm}T_{jm}\Yrjm{2}{jm} \cdot(\Er\odot\Er)
= r \sum_{jm}T_{jm}\, \sNrqj{0}{2}{0}{j}\, \syjm{0}{jm}
= r \sum_{jm} \sqrt{\tfrac{8\pi}{(j+3)!!(2-j)!!}}\, T_{jm}\, \syjm{0}{jm}\ ,
\notag \\
V_\pm &=& r \sum_{jm}T_{jm}\Yrjm{2}{jm} \cdot (\Epm\odot\Er)
= r \sum_{m}T_{2m}\, \sNrqj{\pm1}{2}{0}{2}\, \syjm{\pm1}{2m}
= \mp r \sum_{m} \sqrt{\tfrac{2\pi}{5}}\,T_{2m}\, \syjm{\pm1}{2m}\ .
\eea
Note that since spin weight is limited by $|s|\leq j$,
$V_\pm$ only include $j=2$ components.

While the tensor $T$ is constant,
its helicity-basis components are not.
They can also be expanded in spin-weighted spherical harmonics:
\bea
T_{rr} &=& \sum_{jm} T_{jm}\Yrjm{2}{jm} \cdot(\Er\odot\Er)
= \sum_{jm} T_{jm}\, \sNrqj{0}{2}{0}{j}\, \syjm{0}{jm}
= \sum_{jm}\sqrt{\tfrac{8\pi}{(j+3)!!(2-j)!!}}\, T_{jm}\, \syjm{0}{jm}\ ,
\notag \\
T_{+-} &=& \sum_{jm} T_{jm}\Yrjm{2}{jm} \cdot(\Ep\odot\Em)
= \sum_{jm} T_{jm}\, \sNrqj{0}{2}{1}{j}\, \syjm{0}{jm}
= \sum_{jm}\sqrt{\tfrac{\pi}{2(j+3)!!(2-j)!!}}\,(4-j-j^2)\, T_{jm}\, \syjm{0}{jm}\ ,
\notag \\
T_{r\pm} &=& \sum_{jm} T_{jm}\Yrjm{2}{jm} \cdot(\Er\odot\Epm)
= \sum_{m} T_{2m}\, \sNrqj{\pm1}{2}{0}{2}\, \syjm{\pm1}{2m}
= \mp \sum_{m} \sqrt{\tfrac{2\pi}{5}}\, T_{2m}\, \syjm{\pm1}{2m}\ ,
\notag \\
T_{\pm\pm} &=& \sum_{jm} T_{jm}\Yrjm{2}{jm} \cdot(\Epm\odot\Epm)
= \sum_{m} T_{m}\, \sNrqj{\pm2}{2}{0}{2}\, \syjm{\pm2}{2m}
= \sum_{m} \sqrt{\tfrac{4\pi}{5}} \, T_{2m}\, \syjm{\pm2}{2m}\ .
\eea
Again, only $j=2$ contributes
to the $s=\pm1$ and $s=\pm2$ components.

Notice that any constant symmetric tensor $T$
generates a set of functions
with spins ranging from zero to its rank.
The coefficients in the spherical-harmonic
expansions of all these functions are related
and differ by $\sNrqj{s}{\r}{\qpm}{j}$ factors.
Also notice that we can construct the tensor $T$
given the spherical-harmonic expansions
for any one of the functions in the set.
For example, suppose we were given
a scalar function $f(\mbf r) = \sum_{jm} r^j f_{jm}\yjm{jm}(\N)$
with known coefficients $f_{jm}$.
Using Eq.\ \rf{Y0} or Eq.\ \rf{Ysr},
we can write $f$
in terms of the traceless spherical-harmonics tensors:
\beq
f(\mbf r)
= \sum_{jm}\,
\frac{f_{jm}}{\sNrqj{0}{j}{0}{j}}\,
\Yjm{jm} \cdot \mbf r^{\odot j} \ .
\eeq
This function is the scalar in the
set of tensor-valued functions generated
by the constant tensor 
$T= \sum_{jm} \frac{f_{jm}}{\sNrqj{0}{j}{0}{j}}\,\Yjm{jm}$.

The tensor decomposition of a function
can also be used to quickly calculate derivatives
of a function.
For example,
the gradient of the scalar function $f$ is the vector
\beq
\mbf\nabla f
= \sum_{jm}
\frac{j f_{jm}\,}{\sNrqj{0}{j}{0}{j}}\,
\Yjm{jm} \cdot \mbf r^{\odot (j-1)} \ .
\eeq
The helicity-basis components of the gradient
have the spherical-harmonic expansions
\bea
\Er\cdot \mbf\nabla f
&=& \sum_{jm}
r^{j-1}\, j\, f_{jm}\, \yjm{jm}(\N)\ ,
\notag\\
\Epm\cdot\mbf\nabla f
&=& \sum_{jm}
r^{j-1}j\, f_{jm}\, \frac{\sNrqj{\pm1}{j}{0}{j}}{\sNrqj{0}{j}{0}{j}}\,
\syjm{\pm1}{jm}(\N)\ .
\eea
Vector-calculus operations like this can also be formulated
in terms of the spin-weight ladder operators $J_\pm$ \cite{km09}.

Finally, we note that while the methods
developed here relate the spherical harmonics
to symmetric tensors
they can be applied to other tensors.
Any tensor can be split
into symmetric tensors
with the aid of Young symmetrizers,
which are reviewed in Appendix \ref{young-appendix}.
As an example,
consider an arbitrary rank-3 tensor $T$.
A Young decomposition of
the tensor reveals that it
can be written as
\beq
T = T_S + T_A + T_1 + T_2 \ ,
\eeq
where $T_S^{abc} = \frac{1}{6} T^{(abc)}$
is totally symmetric,
and $T_A^{abc} = \frac{1}{6}T^{[abc]}$
is totally antisymmetric.
There are two mixed-symmetry pieces,
$T_1^{abc} = \frac13\big(T^{abc}+T^{cba}-T^{bac}-T^{cab}\big)$,
which  is antisymmetric in the first two indices,
and
$T_2^{abc} = \frac13\big(T^{abc}+T^{bac}-T^{cba}-T^{bca}\big)$,
which is antisymmetric under interchange of
the first index and last index.
The symmetric part $T_S$ contains 10 independent components,
the antisymmetric $T_A$ has 1 independent component,
and the mixed-symmetry parts have 8 independent components each.

The formalism can be immediately applied to the symmetric part $T_S$.
The antisymmetric part can be written as $T_A^{abc} = \frac{1}{3!}\ep^{abc} f$,
where $\ep^{abc}$ is the antisymmetric Levi-Civita tensor,
and $f = \ep_{abc} (T_A)^{abc}$ is a scalar.
The mixed symmetry pieces can be written as
\bea
(T_1)^{abc} &=& \ep^{ab}{}_d (t_1)^{cd} + (v_1)^{[a}g^{b]c} \ ,
\notag \\
(T_2)^{abc} &=& \ep^{ac}{}_d (t_2)^{bd} + (v_2)^{[a}g^{c]b} \ ,
\eea
where $t_1$ and $t_2$ are symmetric and traceless.
This shows that any rank-3 tensor $T$
can be split into a rank-3 symmetric tensor,
two traceless rank-2 symmetric tensors,
two vectors, and a scalar.
All of these symmetric tensors can be expanded in
spherical harmonics using the above techniques.

\section{Application}
\label{application}

We now turn to a physics application.
We use the spherical-harmonic tensors
to calculate the leading-order effects
of violations of Lorentz invariance
in the gravitational potential $U$,
providing an alternative to the approach
currently found in the literature \cite{newt}.
For experimental tests of Lorentz violation
in Newtonian gravity, see Ref.\ \cite{exps}.

Recent decades have seen renewed interest 
in challenging Lorentz invariance.
These efforts are motivated, in part,
by the observation that Lorentz invariance may be broken
in theories of quantum gravity \cite{strings}.
They were also spurred by the development of the
Standard-Model Extension (SME),
a theoretical framework providing a general description
of all realistic Lorentz violation
in particles and in gravity \cite{sme}.
The SME has served as the theoretical foundation
for hundreds of searches for Lorentz violation \cite{tables}.

The linearized limit of the gauge-invariant
gravitational sector of
the SME is given by the Lagrange density \cite{gw1}
\beq
\cl
= \tfrac14 \ep^{\mu\rh\al\ka} \ep^{\nu\si\be\la} \et_\kl h_\mn \prt_\al\prt_\be h_\rs
+ \sum_d \tfrac14 h_\mn 
\K^{(d)\mu\nu\rh\si\al_1\ldots\al_{d-2}}
\prt_{\al_1}\ldots \prt_{\al_{d-2}}
h_\rs \ ,
\label{lag}
\eeq
where $h_\mn$ is the deviation
of the spacetime metric
from the constant Minkowski metric $\et_\mn$.
The first term in Eq.\ \rf{lag}
is the usual linearized Einstein-Hilbert lagrangian,
which describes conventional gravity in the weak-field limit.
The remaining parts include
all possible Lorentz-violating terms
that are quadratic in $h_\mn$,
translationally invariant,
and invariant under the usual gauge transformation,
$h_\mn \rightarrow h_\mn + \prt_{(\mu} \xi_{\nu)}$.
The Lorentz violation is controlled by
the $\K^{(d)}$ spacetime-tensor coefficients.
A Young decomposition splits
the tensors into three classes
of coefficients for Lorentz violation,
$\s^{(d)}$, $\q^{(d)}$, and $\k^{(d)}$,
with the symmetries given in Table 1 of Ref.\ \cite{gw1}.
The $\s^{(d)}$ coefficients are nonzero for even $d\geq4$,
the $\q^{(d)}$ are nonzero for odd $d\geq5$,
and the $\k^{(d)}$ are nonzero for even $d\geq6$.

The modified equations of motion
arising from Eq.\ \rf{lag} provide
a test theory for studies of Lorentz violation
in gravitational waves \cite{gw1,gw2}
and in Newtonian gravity.
Assuming a static mass distribution $\rh(\mbf x)$,
it can be shown that the Lorentz-violating
contributions to the Newtonian potential 
are given by \cite{newt}
\beq
\de U
= \tfrac14 \sum_{d}
\K^{(d)\mu\mu\nu\nu a_1a_2\ldots a_{d-2}}
\prt_{a_1}\prt_{a_2}\ldots \prt_{a_{d-2}} \ch \ ,
\label{dU}
\eeq
where $\prt_a = \prt/\prt x^a$ are spatial derivatives,
and $d$ is now restricted to even values.
The ``superpotential'' $\ch$ is defined as
\beq
\ch(\mbf x)
= -G_N \int d^3x'\, |\mbf x-\mbf x'|\, \rh(\mbf x')\ ,
\label{chi}
\eeq
where $G_N$ is Newton's constant.
Both the conventional potential
$U = -\frac12 \nabla^2 \ch = G_N
\int d^3x'\, \rh(\mbf x')/|\mbf x-\mbf x'|$
and the Lorentz-violating potential $\de U$
can by found by taking derivatives of $\ch$.

Using the above equations,
one can calculate the effects of Lorentz violation
on gravity from a mass distribution $\rh$,
and experimental constraints can be placed on
the $\K^{(d)\mu\mu\nu\nu a_1a_2\ldots a_{d-2}}$
coefficient combinations.
This is commonly done by searching for variations
in an experimental signal while rotating the apparatus.
In an inertial frame,
these rotations change the mass distribution $\rh$
and the superpotential $\ch$ but not the $\K^{(d)}$ coefficients.
Alternatively,
we can work in a noninertial apparatus-fixed frame
in which $\ch$ is constant but the $\K^{(d)}$ coefficients rotate.
In either case,
the potential $\de U$ varies with these rotations.
The frames commonly used in these types of experiments
and the rotations relating them
are discussed in Appendix \ref{harmonics-appendix}.

The significant role played by rotations
in tests of Lorentz invariance prompts
an angular-momentum decomposition of Eq.\ \rf{dU}.
In Ref.\ \cite{newt},
this is done by switching to momentum space,
which results in the replacement $\prt_a \rightarrow ip_a$.
The $p$-space potential $\de U(\mbf p)$
is then expanded in spherical harmonics.
Using a Fourier transform to switch back to position space,
one finds that the Lorentz-violating potential takes the from
\beq
\de U(\mbf x) = G_N \sum_{djm} \kNdjm{d}{jm}
\int d^3x'\, \frac{\yjm{jm}(\mbf\rhat)\,
  \rh(\mbf x')}{|\mbf r|^{d-3}} \ ,
\label{dUI}
\eeq
where the sum is restricted to
even $j$ and $d = j+2, j+4$.
The vector $\mbf r = \mbf x-\mbf x'$
is the position relative to the source,
and $\mbf\rhat = \mbf r/|\mbf r|$ is the direction.
The spherical Newton coefficients for Lorentz violation $\kNdjm{d}{jm}$
are the linear combinations of components of $\K^{(d)}$
that affect Newtonian gravity at leading order.
The tools developed in this work provide for an alternative derivation
and can be used to find the relationship between
the spherical Newton coefficients
and the coefficient tensors that appear in
the Lagrange density.

We first expand in spherical-harmonic tensors,
\beq
\K^{(d)\mu\mu\nu\nu a_1a_2\ldots a_{d-2}}
= \sum_{jm} \K^{(d)}_{jm}\,
\big(\Yrjm{d-2}{jm}\big)^{a_1a_2\ldots a_{d-2}} \ ,
\label{K1}
\eeq
where
\beq
\K^{(d)}_{jm} = \K^{(d)\mu\mu\nu\nu a_1a_2\ldots a_{d-2}}
\big(\Yrjmconj{(d-2)}{jm}\big)_{a_1a_2\ldots a_{d-2}}\ .
\label{K2}
\eeq
The Lorentz-violating potential can then be written
\beq
\de U = \tfrac14 \sum_{djm} 
\sqrt{\tfrac{(d-2)!(2j+1)!!}{j!(d+j-1)!!(d-j-2)!!}}\
\K^{(d)}_{jm}\,
\big(\Yjm{jm}\big)^{a_1a_2\ldots a_j} \nabla^{d-j-2}
\prt_{a_1}\prt_{a_2}\ldots \prt_{a_j} \ch \ ,
\label{dU1}
\eeq
where $\nabla^2 = \prt_a\prt^a$ is the laplacian.
Since $\nabla^4 \ch = -2 \nabla^2 U  = 0$
outside the mass distribution,
only $d=j+2$ and $d=j+4$ contribute,
giving the restriction on $d$ described above.
Next consider spatial-derivative operators $\prt_a$
acting on the $|\mbf x-\mbf x'| = |\mbf r|$ appearing
in the superpotential \rf{chi}.
Two derivatives give
$\prt_{a_1} \prt_{a_2}|\mbf r|  = g_{a_1a_2}/|\mbf r|
- r_{a_1}r_{a_2}/|\mbf r|^3$.
The term involving the metric will not contribute since
$\Yjm{jm}$ is traceless.
Similar irrelevant terms result when taking higher derivatives.
After a short calculation, this results in
\beq
\de U  = 
\tfrac14 G_N \sum_{djm} 
\xi^d_j\,
(2j-3)!!
\sqrt{\tfrac{(d-2)!(2j+1)!!}{j!(d+j-1)!!(d-j-2)!!}}\, 
\K^{(d)}_{jm}\,
\big(\Yjm{jm}\big)^{a_1a_2\ldots a_j}
\int d^3x'\,  \frac{r_{a_1} \ldots r_{a_j}}{|\mbf r|^{d+j-3}}
\rh(\mbf x')\ ,
\label{dU2}
\eeq
where $\xi^d_j = 1$ when $d=j+2$,
$\xi^d_j = 2-4j$ when $d=j+4$,
and $\xi^d_j = 0$ otherwise.
Using Eq.\ \rf{Y0},
the Lorentz-violating potential reduces to Eq.\ \rf{dUI},
where 
\beq
\kNdjm{d}{jm} = 
\tfrac14 \xi^d_j\,
(2j-3)!!
\sqrt{\tfrac{4\pi (d-2)!}{(d+j-1)!!(d-j-2)!!}}\, 
\K^{(d)}_{jm}\ .
\label{kN}
\eeq
Combined with Eq.\ \rf{K2},
this gives the relationship between the
Newton coefficients 
and the coefficient tensors in the Lagrange density
of the theory.

Finding $\de U$ using Eq.\ \rf{dUI}
requires calculating a different integral
for each $\kNdjm{d}{jm}$ coefficient.
This ``many-integrals'' approach may be
computationally expensive.
The spherical-harmonic tensors lead to a
``many-derivatives'' alternative,
\beq
\de U = \tfrac14 \sum_{djm} \K^{(d)}_{jm}\,
\big(\Yrjm{d-2}{jm}\big)^{a_1a_2\ldots a_{d-2}}
\prt_{a_1}\prt_{a_2}\ldots \prt_{a_{d-2}} \ch \ ,
\eeq
which may be easier to compute.
Only a single integral is required
in order to calculate the superpotential,
and the effects of Lorentz violation are then
found by taking its derivatives.

\section{Summary}

In this work,
we construct an orthonormal set of
symmetric spherical-harmonic tensors.
The most general versions are the rank-$\r$ tensors
$\Yrjm{\r}{jm}$ given in Eq.\ \rf{Yrjm}.
Their connection to spherical harmonics $\yjm{jm}$
is given in  Eq.\ \rf{Y0r} and
to spin-weighted spherical harmonics $\syjm{s}{jm}$
in Eq.\ \rf{Ysr}.
In the case in which the rank $\r = j$,
the tensors $\Yjm{jm} = \Yrjm{j}{jm}$ are traceless
and reduce to Eq.\ \rf{Yjm}.
Equation \rf{Y0} gives the relation between the
$\Yjm{jm}$ tensors and
the scalar $s=0$ spherical harmonics $\yjm{jm}$.
The $\Yrjm{\r}{jm}$ are constant tensors
with spin eigenvalues $J^2=j(j+1)$ and $J_z=m$
and
form an angular-momentum basis for symmetric rank-$\r$ tensors.
Any constant rank-$\r$ tensor $T$ can be expanded in $\Yrjm{\r}{jm}$,
providing a full angular-momentum and trace decomposition.

Section \ref{examples} contains
several examples illustrating how the $\Yrjm{\r}{jm}$
can be used to expand tensors and
tensor-valued functions in $\Yrjm{\r}{jm}$
and how they are connected to spherical-harmonic expansions.
An application of the formalism involving the study of violations
of Lorentz invariance is discussed in Sec.\ \ref{application}.
The leading-order effects of potential Lorentz violation
in Newtonian gravity are formulated in terms
of derivatives of a gravitational superpotential
using the $\Yjm{jm}$ tensors,
providing a ``many-derivatives'' alternative
to the ``many-integrals'' approach that is currently
found in the literature.

\section*{Acknowledgments}

This work was supported in part
by the William and Linda Frost Fund
and by the United States National Science Foundation 
under grant number PHY-1819412.

\appendix

\section{Spin-weighted spherical harmonics}
\label{harmonics-appendix}

Spin-weighted spherical harmonics
are a form of tensor spherical harmonics
and provide an angular-momentum decomposition
for functions with both nonzero spin
and orbital angular momentum.
This appendix briefly reviews
spin weight and spin-weighted spherical harmonics. 
A more detailed discussion can be found
in Refs.\ \cite{sYjm1,sYjm2,sYjm3,km09}.

To understand spin weight,
first consider the various sets
of compatible angular-momentum operators.
The physics literature typically focuses on
the product-space basis,
comprised of eigenfunctions of the operators $\{S^2,L^2,S_z,L_z\}$,
and the total-angular-momentum basis,
given by the eigenfunctions of $\{S^2,L^2,J^2,J_z\}$.
However, a third set exists,
$\{S^2,J^2,J_z,J_r\}$,
where the helicity $J_r = \N\cdot\mbf J = \N\cdot\mbf S$
is the component of the total angular momentum
or spin angular momentum along the $\N$ direction.
The spin-weighted spherical harmonics $\syjm{s}{jm}(\N)$
are eigenfunctions of $J^2$, $J_z$ and $J_r$.
By convention,
the spin weight $s$ is defined so that it is the eigenvalue of $-J_r$,
implying spin weight is the opposite of helicity.
It is limited by $-j\leq s\leq j$
since it is a component of the total angular momentum.
The usual harmonics correspond to the $s=0$ case,
$\yjm{jm} = \syjm{0}{jm}$.
More generally,
the spin-weighted spherical harmonics $\syjm{s}{jm}$
form an orthonormal basis for spin-weighted functions
and provide for the angular-momentum expansion
of higher-rank tensor functions.

A function $f(\N)$ is said to have spin-weight $s$
if it transforms according to
$f\rightarrow e^{-is\al} f$
under an active rotation about $\N$ by angle $\al$.
These rotations are generated by the helicity operator $J_r=S_r$.
This only depends on spin since orbital angular momentum $\mbf L$
transforms the argument of a function,
and $\N$ is invariant under these rotations.
Spin $\mbf S$ accounts for the directionality of an object.
Spin-weighted functions with $s\neq 0$
change under rotations generated by $S_r$,
implying they are necessarily directional or tensoral in nature.
The reverse is also true.
Tensors of nonzero rank have spin weight.
More specifically,
the components of a tensor in the helicity basis
$\{\Er, \Ep, \Em\}$
are spin-weighted functions \cite{km09}.

The spin weight of a helicity-basis component of a tensor
is determined by the number
of $+$ or $-$ indices.
Each lowered $\pm$ index or raised $\mp$
index contributes $\pm1$ to the spin weight.
For example,
a rank-3 tensor has 
six $s=1$ components,
$T_{+rr}$, $T_{r+r}$, $T_{rr+}$,
$T_{++-}$, $T_{+-+}$, and $T_{-++}$,
each of which can be expanded in
the $\syjm{1}{jm}$ spherical harmonics.
In total,
the helicity-basis components of $T$
give 27 spin-weighted functions:
one for each of $s=\pm 3$,
three for each of $s=\pm 2$,
six for each of $s=\pm 1$,
and seven with $s=0$.
In general,
the spin-weight of the components of a rank-$\r$ tensor
is limited by $-\r\leq s\leq \r$.

Many of the usual spherical-harmonic identities
can be extended to the spin-weighted harmonics.
Harmonics of equal spin weight are orthonormal,
\beq
\int \syjm{s}{jm}^*\, 
\syjm{s}{j'm'} \, 
\sin\th d\th d\ph = 
\de_{jj'}\de_{mm'} \ ,
\eeq
and satisfy the completeness relations
\beq
\sum_{jm} 
\syjm{s}{jm}^*(\th,\ph)\, 
\syjm{s}{jm}(\th',\ph')
= \frac{\de(\th - \th')\de(\ph-\ph')}{\sin\th}\ .
\label{compl}
\eeq
Assuming a Condon-Shortley phase,
they obey the complex-conjugation rule
\beq
\syjm{s}{jm}^*=(-1)^{s+m}\, \syjm{-s}{j(-m)}\ .
\label{conj}
\eeq
They also obey the parity relation
\beq
\syjm{s}{jm}(-\N) =(-1)^j\, \syjm{-s}{jm}(\N) \ .
\label{parity}
\eeq
The product of two harmonics is
\beq
\syjm{s_1}{j_1m_1}~\syjm{s_2}{j_2m_2} = \sum_{s_3j_3m_3} 
\sqrt{\tfrac{(2j_1+1)(2j_2+1)}{4\pi(2j_3+1)}}
\cg{j_1j_2(-s_1)(-s_2)}{j_3 (-s_3)}
\cg{j_1j_2m_1m_2}{j_3 m_3}\,
\syjm{s_3}{j_3m_3}\ ,
\label{Yproduct}
\eeq
where $\cg{j_1j_2m_1m_2}{j_3m_3}$ are Clebsch-Gordan coefficients.

The spin-weighted harmonics are eigenfunctions
of the square of the total angular momentum $J^2$,
the $z$-component of the total angular momentum $J_z$,
and the helicity $J_r$,
with eigenvalues $J^2=j(j+1)$, $J_z=m$, and $J_r = -s$.
The spin-weighted harmonics transform relatively simply under rotations
since they are generated by angular momentum $\vec J$.
All three components of $\vec J$ commute with both $J^2$ and $J_r$,
implying quantum numbers $j$ and $s$ are invariant under rotations,
and rotations only mix harmonics with different $m$ values,
leaving $j$-$s$ subspaces invariant.
The mixing is characterized by the Wigner matrices,
defined through
\beq
D^{(j)}_{mm'}(\al,\be,\ga) = 
\int \syjm{s}{jm}^* e^{-i\al J_z}e^{-i\be J_y}e^{-i\ga J_z}\, \syjm{s}{jm'}\,
\sin\th d\th d\ph\ ,
\label{Wigner}
\eeq
where $\al$, $\be$, and $\ga$ are Euler angles.
Note that Eq.\ \rf{Wigner} assumes $|s|\leq j$,
but is otherwise independent of the the spin weight $s$.
The above employs a $z$-$y$-$z$ rotation convention,
which is advantageous since the spherical harmonics
are eigenfunctions for $z$ rotations,
leading to simple phases for two of the Euler angles,
\beq
D^{(j)}_{mm'}(\al,\be,\ga) = 
e^{-i\al m} e^{-i\ga m'}\, d^{(j)}_{mm'}(\be)\ , 
\label{wigner}
\eeq
where $d^{(j)}_{mm'}(\be)=D^{(j)}_{mm'}(0,\be,0)$
are the little Wigner matrices.

Operating on the components of a tensor,
$e^{-i\al J_z}e^{-i\be J_y}e^{-i\ga J_z}$
rotates tensor components by $\ga$ about the $z$ axis,
then by $\be$ about the $y$ axis,
and finally by $\al$ about the $z$ axis.
Interpreting this as a passive transformation,
this corresponds to rotating the coordinate axes
by $-\ga$ about the $z$ axis,
then $-\be$ about the rotated $y$ axis,
and then by $-\al$ about the new $z$ axis.
The cartesian components of the old frame $\{x,y,z\}$
and the rotated frame $\{x',y',z'\}$ are related through
\beq
\begin{pmatrix}
  x'\\ y'\\ z'
\end{pmatrix}
=
\begin{pmatrix}
  \cos\al&-\sin\al&0\\
  \sin\al&\cos\al&0\\
  0&0&1
\end{pmatrix}
\begin{pmatrix}
  \cos\be& 0&\sin\be\\
  0&1&0\\
  -\sin\be& 0&\cos\be
\end{pmatrix}
\begin{pmatrix}
  \cos\ga&-\sin\ga&0\\
  \sin\ga&\cos\ga&0\\
  0&0&1
\end{pmatrix}
\begin{pmatrix}
  x\\ y\\ z
\end{pmatrix} \ .
\eeq
The basis vectors $\{\Ex,\Ey,\Ez\}$
and the rotated-basis vectors $\{\Ex',\Ey',\Ez'\}$
also obey this relation.

We can transform the spherical-harmonic expansion
of a function using the Wigner matrices.
A spin-weighted function
$f(\N) = \sum_{jm} f_{jm}\, \syjm{s}{jm}(\N)$
rotates according to 
$f'(\N) = \sum_{jm} f'_{jm}\, \syjm{s}{jm}(\N)
= e^{-i\al J_z}e^{-i\be J_y}e^{-i\ga J_z} f(\N)$,
giving rotated expansion coefficients
\beq
f'_{jm} = \sum_{m'} D^{(j)}_{mm'}(\al,\be,\ga) f_{jm'} \ .
\eeq
Consider, for example,
rotations of a laboratory
due to the daily rotation of the Earth.
Standard reference frames appear
in the literature to account for this rotation \cite{km09,tables}.
A nonrotating Sun-centered frame
is defined so that the $\mbf{\hat Z}$ axis points
along the Earth's rotation axis
and $\mbf{\hat X}$ and $\mbf{\hat Y}$ lie in
the equatorial plane with right ascension
$0^\circ$ and $90^\circ$, respectively.
A rotating laboratory-fixed frame
is defined with $\mbf{\hat z}$ pointing up
and $\mbf{\hat x}$ and $\mbf{\hat y}$ horizontal
with $\mbf{\hat x}$ at an angle $\vp$ measured east of south.
The coordinates in these two frames are related through
\beq
\begin{pmatrix}
  X\\ Y\\ Z
\end{pmatrix}
=
\begin{pmatrix}
  \cos\al&-\sin\al&0\\
  \sin\al&\cos\al&0\\
  0&0&1
\end{pmatrix}
\begin{pmatrix}
  \cos\ch& 0&\sin\ch\\
  0&1&0\\
  -\sin\ch& 0&\cos\ch
\end{pmatrix}
\begin{pmatrix}
  \cos\vp&-\sin\vp&0\\
  \sin\vp&\cos\vp&0\\
  0&0&1
\end{pmatrix}
\begin{pmatrix}
  x\\ y\\ z
\end{pmatrix} \ ,
\eeq
where $\ch$ is the colatitude of the laboratory,
and $\al$ is the right ascension of the laboratory zenith.
The spherical-expansion coefficients
in the two frames are related through
\beq
f^\text{Sun}_{jm}
= \sum_{m'} D^{(j)}_{mm'}(\al,\ch,\vp) f^\text{lab}_{jm'} \ ,
\qquad
f^\text{lab}_{jm}
= \sum_{m'} D^{(j)}_{mm'}(-\vp,-\ch,-\al) f^\text{Sun}_{jm'} \ .
\eeq
Note that $\al$ increases at Earth's sidereal rate
$\approx 2\pi/\text{23 hr 56 min}$
due to the daily rotation of the Earth.
It is common for experiments to be placed on horizontal turntables
so that the angle $\vp$ varies as well.

\section{Young symmetrizers}
\label{young-appendix}

This appendix provides a brief overview of Young symmetrizers
and their use in the symmetry decomposition of tensors.
For an in-depth treatment, see, e.g., Refs.\ \cite{group1,group2}.
The splitting of a rank-2 tensor $T^{ab}$ into its symmetric part
$T_\text{S}^{ab} = \frac12 T^{(ab)} = \frac12 (T^{ab}+T^{ba})$
and its antisymmetric part
$T_\text{A}^{ab} = \frac12 T^{[ab]} = \frac12 (T^{ab}-T^{ba})$
is a simple example of Young symmetrization.
Each of these pieces transforms under a different
irreducible representation of the general linear group $GL_n$,
meaning they do not mix under linear transformations
of the underlying $n$-dimensional vector space.
Note that a trace decomposition provides additional
reduction for orthogonal subgroups,
such as rotations or Lorentz transformations.
For example, the trace decomposition
of a rank-2 tensor yields
$T^{ab} = \smash{\overline T}_\text{S}^{ab} + T_\text{A}^{ab}
+ \frac{1}{n} g^{ab} T_\text{tr}$,
where $T_\text{tr} = T^{ab}g_{ab}$ is the trace,
and $\smash{\overline T}_\text{S}^{ab} = T_\text{S}^{ab}
- \frac{1}{n} g^{ab} T_\text{tr}$
is the traceless symmetric part.

Tensors $T^{a_1\ldots a_\r}$ with larger rank $\r>2$
contain additional mixed-symmetry parts.
These can be constructed using Young symmetry projections,
each of which can be represented graphically as a Young tableau.
The construction of a tableau starts by selecting
indices from $T^{a_1\ldots a_\r}$ in any order.
Here we take them in the order they appear
and start by drawing a single-box tableau $\BT{1}{1}\BX{a_1}\ET$
for the first index.
We then combine this with the next index to produce
the two-box tableaux:
\beq
\BT{1}{1}\BX{a_1}\ET \otimes
\BT{1}{1}\BX{a_2}\ET
=
\BT{2}{1}\BX{a_1}\BX{a_2}\ET
\oplus
\BT{1}{2}\BX{a_1}\ROW\BX{a_2}\ET \ .
\eeq
Each subsequent index creates a new row
or extends an existing row.
For example,
the three-box tableaux are
\beq
\Big(\
\BT{2}{1}\BX{a_1}\BX{a_2}\ET
\oplus
\BT{1}{2}\BX{a_1}\ROW\BX{a_2}\ET
\ \Big)
\otimes\,
\BT{1}{1}\BX{a_3}\ET
=
\BT{3}{1}\BX{a_1}\BX{a_2}\BX{a_3}\ET
\,\oplus\,
\BT{2}{2}\BX{a_1}\BX{a_2}\ROW\BX{a_3}\ET
\,\oplus\,
\BT{2}{2}\BX{a_1}\BX{a_3}\ROW\BX{a_2}\ET
\,\oplus\,
\BT{1}{3}\BX{a_1}\ROW\BX{a_2}\ROW\BX{a_3}\ET \ .
\eeq
Adding another index gives
\beq
\BT{4}{1}\BX{a_1}\BX{a_2}\BX{a_3}\BX{a_4}\ET
\,\oplus\,
\BT{1}{4}\BX{a_1}\ROW\BX{a_2}\ROW\BX{a_3}\ROW\BX{a_4}\ET
\,\oplus\,
\BT{3}{2}\BX{a_1}\BX{a_2}\BX{a_3}\ROW\BX{a_4}\ET
\,\oplus\,
\BT{3}{2}\BX{a_1}\BX{a_2}\BX{a_4}\ROW\BX{a_3}\ET
\,\oplus\,
\BT{3}{2}\BX{a_1}\BX{a_3}\BX{a_4}\ROW\BX{a_2}\ET
\,\oplus\,
\BT{2}{3}\BX{a_1}\BX{a_2}\ROW\BX{a_3}\ROW\BX{a_4}\ET
\,\oplus\,
\BT{2}{3}\BX{a_1}\BX{a_3}\ROW\BX{a_2}\ROW\BX{a_4}\ET
\,\oplus\,
\BT{2}{3}\BX{a_1}\BX{a_4}\ROW\BX{a_2}\ROW\BX{a_3}\ET
\,\oplus\,
\BT{2}{2}\BX{a_1}\BX{a_2}\ROW\BX{a_3}\BX{a_4}\ET
\,\oplus\,
\BT{2}{2}\BX{a_1}\BX{a_3}\ROW\BX{a_2}\BX{a_4}\ET
 \ .
\eeq
One continues in this fashion for all $\r$ indices.
Note that this produces a smaller number of shapes
known as Young diagrams
and that tableaux with the same diagram
are related through permutations of indices.

Young symmetrizers are combinations
of symmetrization operators $\S$
and antisymmetrization operators $\A$,
defined so that
$\S(a_2a_3a_4) T^{a_1a_2a_3a_4a_5\ldots}
= T^{a_1(a_2a_3a_4)a_5\ldots}$
and
$\A(a_2a_3a_4) T^{a_1a_2a_3a_4a_5\ldots}
= T^{a_1[a_2a_3a_4]a_5\ldots}$,
for example.
The Young symmetrizer $\P$ associated with a given tableau
is constructed by first symmetrizing on indices in each row
then antisymmetrizing on indices in each column.
For example,
the Young symmetrizers for the $\r=4$ tableaux are
\begin{alignat}{4}
\BT{4}{1}\BX{a_1}\BX{a_2}\BX{a_3}\BX{a_4}\ET
:&\ & \P_1 &=  \C_1\, \S(a_1a_2a_3a_4) \ ,
&\qquad
\BT{2}{3}\BX{a_1}\BX{a_2}\ROW\BX{a_3}\ROW\BX{a_4}\ET
:&\ & \P_6 &=  \C_6\, \A(a_1a_3a_4)\S(a_1a_2) \ ,
\notag\\
\BT{1}{4}\BX{a_1}\ROW\BX{a_2}\ROW\BX{a_3}\ROW\BX{a_4}\ET
:&\ & \P_2 &=  \C_2\, \A(a_1a_2a_3a_4) \ ,
&\qquad
\BT{2}{3}\BX{a_1}\BX{a_3}\ROW\BX{a_2}\ROW\BX{a_4}\ET
:&\ & \P_7 &=  \C_7\, \A(a_1a_2a_4)\S(a_1a_3) \ ,
\notag\\
\BT{3}{2}\BX{a_1}\BX{a_2}\BX{a_3}\ROW\BX{a_4}\ET
:&\ & \P_3 &=  \C_3\, \A(a_1a_4)\S(a_1a_2a_3) \ ,
&\qquad
\BT{2}{3}\BX{a_1}\BX{a_4}\ROW\BX{a_2}\ROW\BX{a_3}\ET
:&\ & \P_8 &=  \C_8\, \A(a_1a_2a_3)\S(a_1a_4) \ ,
\notag\\
\BT{3}{2}\BX{a_1}\BX{a_2}\BX{a_4}\ROW\BX{a_3}\ET
:&\ & \P_4 &=  \C_4\, \A(a_1a_3)\S(a_1a_2a_4) \ ,
&\qquad
\BT{2}{2}\BX{a_1}\BX{a_2}\ROW\BX{a_3}\BX{a_4}\ET
:&\ & \P_9 &=  \C_9\, \A(a_1a_3)\A(a_2a_4)\S(a_1a_2)\S(a_3a_4) \ ,
\notag\\
\BT{3}{2}\BX{a_1}\BX{a_3}\BX{a_4}\ROW\BX{a_2}\ET
:&\ & \P_5 &=  \C_5\, \A(a_1a_2)\S(a_1a_3a_4) \ ,
&\qquad
\BT{2}{2}\BX{a_1}\BX{a_3}\ROW\BX{a_2}\BX{a_4}\ET
:&\ & \P_{10} &=  \C_{10}\, \A(a_1a_2)\A(a_3a_4)\S(a_1a_3)\S(a_2a_4) \ .
\end{alignat}
The $\C_\io$ constants are chosen
so that the $\P_\io$ form a complete set of
orthogonal projection operators:
$\sum_\io\P_\io = \text{identity}$,
$\P_\io\P_{\io'} = \P_\io \de_{\io\io'}$.
The projections $T_\io=\P_\io T$ give parts 
of a tensor $T= \sum_\io T_\io$ that transform under
irreducible representations of $GL_n$.
For example, $\P_9$ from above gives the tensor
\begin{align}
T_9^{a_1a_2a_3a_4} = \C_9\big(
&T^{a_1a_2a_3a_4} + T^{a_2a_1a_3a_4}
+ T^{a_1a_2a_4a_3} + T^{a_2a_1a_4a_3}
\notag \\ \ \ +
&T^{a_3a_4a_1a_2} + T^{a_4a_3a_1a_2}
+ T^{a_3a_4a_2a_1} + T^{a_4a_3a_2a_1}
\notag \\ \ \ -
&T^{a_3a_2a_1a_4} - T^{a_2a_3a_1a_4}
- T^{a_3a_2a_4a_1} - T^{a_2a_3a_4a_1}
\notag \\ \ \ -
&T^{a_1a_4a_3a_2} - T^{a_4a_1a_3a_2}
- T^{a_1a_4a_2a_3} - T^{a_4a_1a_2a_3}
\big)\ .
\end{align}
Note $T_1$ and $T_2$ in our example
are the fully symmetric and antisymmetric parts.
The other parts are said to have mixed symmetry.
Also note that the $T_\io$ are antisymmetric in indices
appearing in each column of the tableau.
This implies a $T_\io$ vanishes
if the number of rows in the tableau
exceeds the dimension $n$ of the space.
The $T_\io$  are not symmetric in indices appearing in the rows.
However, different conventions exist,
including ones where $\S$ operations follow $\A$ operations,
leaving the $T_\io$ symmetric in row indices.

The mixed-symmetry $T_\io$ can have complicated symmetries
leading to more subtle features,
which can be uncovered using symmetrizers.
For example,
the antisymmetrization of any three indices of $T_9^{a_1a_2a_3a_4}$
vanishes since
\beq
\A(a_1a_2a_3)\P_9 =
\C_9\, \A(a_1a_2a_3) \A(a_1a_2)\S(a_1a_3a_4)
= 2\C_9\, \A(a_1a_2a_3) \S(a_1a_3a_4)
= 0\ .
\eeq
The same holds for any choice of three indices.

\begin{figure}
  \includegraphics{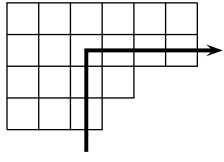}
  \caption{\label{hook}
    Hook for row $r=2$ and column $c=3$
    with hook length $h_{2,3}=6$.}
\end{figure}

The $\C_\io$ normalization constants
and the number $\Num_\io$ of independent components
for a given $T_\io$ can be easily calculated using
the hook lengths $h_{r,c}$ for the diagram \cite{hooks}.
A hook is a path through a Young diagram
moving up through the bottom of
the diagram to the box in row $r$ and column $c$
and then out of the diagram to the right.
The hook length is the number boxes the hook passes through.
An example is shown in Fig.\ \ref{hook}.
The $\C_\io$ normalization constant is
the reciprocal of the product
of all the hook lengths for the diagram,
\beq
\C_\io = \prod_{r,c}\frac{1}{h_{r,c}}\ ,
\eeq
and the number of independent components
in tensor $T_\io$ is given by
\beq
\Num_\io = \prod_{r,c}\frac{n+c-r}{h_{r,c}}\ ,
\eeq
where again $n$ is the dimension of the space.
As an example,
the $\P_9$ above has normalization constant $\C_9 = 1/12$,
and tensor $T_9$ has $\Num_9 = (n+1)n^2(n-1)/12$ independent components.
The independent components
can be taken as those corresponding to semistandard tableau,
i.e., ones where the index values increase as one moves down a column
and do not decrease when moving to the right in a row.
For example,
taking $n=3$ and working in cartesian coordinates $\{x,y,z\}$,
the $\Num_9 = 6$ semistandard tableau for $\P_9$ are
\beq
\BT{2}{2}\BX{a_1}\BX{a_2}\ROW\BX{a_3}\BX{a_4}\ET =
\BT{2}{2}\BX{x}\BX{x}\ROW\BX{y}\BX{y}\ET\ ,\
\BT{2}{2}\BX{x}\BX{x}\ROW\BX{y}\BX{z}\ET\ ,\
\BT{2}{2}\BX{x}\BX{x}\ROW\BX{z}\BX{z}\ET\ ,\
\BT{2}{2}\BX{x}\BX{y}\ROW\BX{y}\BX{z}\ET\ ,\
\BT{2}{2}\BX{x}\BX{y}\ROW\BX{z}\BX{z}\ET\ ,\
\BT{2}{2}\BX{y}\BX{y}\ROW\BX{z}\BX{z}\ET\ .
\eeq
Using the symmetries of $T_9$,
all of its components $T_9^{a_1a_2a_3a_4}$
can be written as linear combinations of
$T_9^{xxyy}$,
$T_9^{xxyz}$,
$T_9^{xxzz}$,
$T_9^{xyyz}$,
$T_9^{xyzz}$, and
$T_9^{yyzz}$.

\end{document}